\documentclass[10pt, conference, compsocconf,fleqn]{IEEEtran} %TWO COLUMN

\usepackage{times}
\usepackage{amsmath}
\usepackage{amssymb,amsbsy,latexsym}
\usepackage{enumerate}
\usepackage{graphicx}
\usepackage{tabularx}
\usepackage{color, colortbl}
\usepackage[aboveskip=1cm]{subfig}

\begin{document}

\title{Side-Channel Attack Resilience through Route Randomisation in Secure Real-Time Networks-on-Chip}

%\author{Leandro Soares Indrusiak, James Harbin and Martha Johanna Sepulveda,\\
%Department of Computer Science,\\University of York,\\York, YO10 5GH, UK}

\author{
\IEEEauthorblockN{Leandro Soares Indrusiak}
\IEEEauthorblockA{Department of Computer Science,\\
University of York, UK.\\
Email: leandro.indrusiak@york.ac.uk} \and
\IEEEauthorblockN{James Harbin }
\IEEEauthorblockA{Department of Computer Science,\\
University of York, UK.\\
Email: james.harbin@york.ac.uk} \and
\IEEEauthorblockN{Martha Johanna Sepulveda}
\IEEEauthorblockA{Institute for Security in Information Technology,\\
Technische Universitaet Muenchen, Germany.\\
Email: johanna.sepulveda@tum.de} }

\maketitle

\begin{abstract}

Security can be seen as an optimisation objective in NoC resource management, and as such poses trade-offs against other objectives such as real-time schedulability. In this paper, we show how to increase NoC resilience against a concrete type of security attack, named side-channel attack, which exploit the correlation between specific non-functional properties (such as packet latencies and routes, in the case of NoCs) to infer the functional behaviour of secure applications. For instance, the transmission of a packet over a given link of the NoC may hint on a cache miss, which can be used by an attacker to guess specific parts of a secret cryptographic key, effectively weakening it. 

We therefore propose packet route randomisation as a mechanism to increase NoC resilience against side-channel attacks, focusing specifically on the potential impact of such
an approach upon hard real-time systems, where schedulability is a vital design requirement. Using an evolutionary optimisation approach, we show how to effectively apply route randomisation in such a way that it can increase NoC security while controlling its impact on hard real-time performance guarantees. Extensive experimental evidence based on analytical and simulation models supports our findings.

\end{abstract}

\section{Introduction}

The design of Network-on-Chip (NoC) interconnects for embedded systems
requires the careful balance of multiple trade-offs. Over the past
decades, a significant amount of work has addressed the trade-offs
between performance and other secondary objectives such as
energy~\cite{Silvano10}, fault-tolerance~\cite{Radetzki13}, and chip
area~\cite{Pestana04}. Less work has addressed such trade-offs in NoCs
with hard real-time constraints, with some inroads towards improving
energy~\cite{Sayuti13} and area efficiency (by optimising buffering in
virtual channels ~\cite{Nikolic13}) while meeting deadlines of all
packets even in the worst-case scenario.

In this paper, we consider NoCs with hard real-time constraints, and
address a novel trade-off that has increasing importance in embedded
systems: security. Because of their key role in interconnecting the
multiple components of an embedded system, NoCs can be seen as a major
security vulnerability. If an attacker can extract information from
the NoC interconnect, they can potentially compromise the security of
the complete embedded system. Therefore, many mechanisms have been
designed to improve NoC security (as reviewed in Section \ref{RelWor})
and many more will certainly be developed in the coming
years. However, most of such mechanisms impose performance overheads,
and therefore can potentially jeopardise the ability of the NoC to
provide real-time guarantees. So we argue in this paper that, just
like in the previously mentioned trade-offs, security can be seen as
an optimisation objective in NoC resource management: designers must
carefully consider the resources they have available to increase NoC
security without sacrificing performance guarantees (which in the case
of hard real-time NoCs will always be the primary objective).

In Section \ref{RelWor}, we review the most relevant NoC security
mechanisms and the types of attacks they aim to prevent, discussing
their performance overheads and resource usage, and highlighting the
need for the contributions of this paper. Then, in Section \ref{Prob}
we provide details on the specific problem we address in this paper,
which is a security mechanism that aims to improve the NoC resilience
to side-channel attacks. Such attacks try to break a secure system by
gathering information from the system's timing behaviour, power
consumption, temperature or electromagnetic emissions.  Just like some
of the related work~\cite{WAS14}~\cite{ESL15}, we address the problem
of side channel attacks by randomising the behaviour of the NoC,
aiming to make it difficult for an attacker to identify patterns and
correlations between the functionality of the system and the timing,
power, temperature and electromagnetic behaviour of the NoC. As
expected, such an approach has a direct impact on NoC resource usage,
and therefore on its real-time guarantees, so in Section \ref{Random}
we identify techniques that support NoC designers in improving NoC
resilience against side-channel attacks while still maintaining full
system schedulability. The paper is closed with extensive experimental
work based on schedulability analysis and simulation in Section
\ref{Eval}, and with a summary of our findings.

\section{Related Work}\label{RelWor}

Multiprocessor embedded systems are target of attacks by means of
malicious hardware or software~\cite{DIG07}. Hardware-based attacks
depend on design-time access to the system, which is then modified in
a way that can be exploited during operation (e.g. by adding hardware
able to leak information by changing chip
temperature~\cite{Iakymchuk11}). Software-based attacks are the most
common cause of security incidents in such types of
systems~\cite{PAP15}, and are carried out by malicious software
installed at design time or after deployment.

NoC-based systems have been shown to be vulnerable to a variety of
attacks, both hardware and software-based. Active NoC attacks, such as
code injection~\cite{DAC15}, malware ~\cite{FIO08} and control
hijacking ~\cite{LEK11}, or passive NoC attacks, such as side-channel
exploitation, can be used to read sensitive communications, modify the
system behaviour or prevent correct NoC operation.  NoCs are
especially vulnerable to side-channel attacks that exploit traffic
interference as timing channels ~\cite{WAN12}~\cite{ESL15}. The shared
nature of NoCs can be exploited by an attacker to obtain sensitive
information. By forcing traffic collision with sensitive packet flows,
an attacker can observe the throughput variations and infer sensitive
data, as shown in ~\cite{WAN12}~\cite{ESL15}~\cite{WAS14}.
 
Security-enhancing mechanisms have been added to NoC platforms to
provide authentication~\cite{SEP12}, access control~\cite{FIO08},
integrity~\cite{ICE14}, and confidentiality services~\cite{REC15}. By
monitoring and controlling the data exchange inside the chip, NoCs can
detect and avoid attacks.

Firewall-based and crypto-based techniques integrated at the network
interface are the most commonly used approaches against active NoC
attacks over the past decade~\cite{FIO08}~\cite{PAS16}. Firewalls
implement authentication, access control and integrity services by
means of traffic matching with a security table. Authorized
transactions are allowed and injected to the NoC, otherwise they are
denied and thus dropped. Crypto-based NoCs implement the
confidentiality service by creating a shared secret among the
sensitive cores and perform the encoded data exchange. While achieving
desirable security enhancements, such approaches have an unpredictable
impact upon the performance of the NoC and thus the overall
system.

Firewalls and crypto-based NoCs are the state-of-the-art in NoC
security, but they are not able to protect the system against passive
NoC attacks. Randomised arbitration~\cite{ESL15}, virtual channel
allocation~\cite{FLO16} and routing~\cite{WAS14} have been
investigated and evaluated as countermeasures against timing
attacks. By randomising the characteristics of sensitive packet flows,
it is possible to break the correlation between the traffic
characteristics (e.g. volume and access patterns) and the sensitive
data thus avoiding information leakage. Among those mechanisms, random
routing has achieved the best levels of security enhancement with the
lowest energy and area overhead~\cite{WAS14}. By spreading sensitive
traffic over the NoC, the spatial distribution makes it harder for
compromised cores or external attackers to gather sufficient
side-channel information to infer correlations with sensitive data.

Similarly to firewalls and crypto-based approaches, the focus of
randomisation approaches is to increase security and none of the works
in the state-of-the-art consider the performance requirements of the
applications. In this paper, we argue that NoCs supporting real-time
applications require a careful balance of a trade-off between security
and performance. In most cases, we envisage that the level of security
will be constrained by the NoC's ability to support attack
countermeasures while at the same time ensuring performance guarantees
to the application.

Thus, the main contributions of this paper are the identification of a
test to evaluate whether performance guarantees can hold under a
specific side-channel attack countermeasure (namely route
randomisation), and a technique that uses that test to better balance
the trade-off between performance guarantees, resource usage and
security.

\section{Problem Description}\label{Prob}

\subsection{Network-on-Chip Architecture}\label{NoC}

While the contribution of this paper can be applied to a large variety
of NoC architectures, we believe it is easier to explain it with the
help of a concrete architecture. Therefore, we open this section with
the description of such architecture, and postpone to Section
\ref{conc} the discussion about the applicability of the proposed
approach to other NoC architectures.

We assume a NoC architecture with a 2D-mesh topology and wormhole
switching protocol, because such features are commonly used in
embedded systems for their simplicity and moderate resource overheads:
\begin{itemize}
  \item In a 2D-mesh topology, every core is connected to a NoC switch
    via a network interface (NI), which is responsible for packetising
    and depacketising data, and controlling the injection of packets
    into the network. We use the term core very loosely, so it can
    mean a processing core, a memory controller, an I/O controller
    (e.g. wireless communication interface) or any other hardware
    resource that requires chip-level communication. The regularity of
    such a topology is attractive because it simplifies packet routing,
    and because it facilitates chip floorplanning, placement and
    routing.

  \item The use of wormhole switching protocols allows packets to be
    gradually sent over the NoC in smaller units called flits. Once a
    flit is received by a switch, it can be forwarded to the next
    switch down the packet route as long as that switch has sufficient
    buffering to hold it. This means that at any given time a packet
    could have its flits temporarily stored by multiple switches, so
    each of them are not required to hold a complete packet, thus
    reducing the overall buffering requirements of the NoC.
\end{itemize}

There is a downside to this choice of topology and switching protocol,
which is the difficulty in predicting packet latencies. Since a packet
can be simultaneously occupying multiple NoC buffers and links, there
is a significant amount of competition for resources throughout the
NoC at all times. The wide variety of interference patterns makes it
hard to predict how long it takes for a packet to reach its
destination. Different resource arbitration policies can make such
predictions more or less difficult, especially in the case of hard
real-time NoCs when an upper-bound worst-case latency is
needed. Previous work has considered NoC arbitration based on packet
priority~\cite{Shi10}, time multiplexing~\cite{Schoeberl07} and round
robin~\cite{Dasari14}, and has devised analytical models that can be
used to find latency upper-bounds for packet flows transmitted over
such NoCs~\cite{Kiasari13}. Any of those approaches could be used in
this paper, and we chose a priority-arbitrated NoC because of its
ability to provide upper-bound latency guarantees that are
customisable to different levels of packet urgency while allowing for
high NoC link utilisation~\cite{Indrusiak14}.

\subsection{Side-channel Attacks and Countermeasures}

In this paper, we aim to improve NoC resilience against side-channel
attacks by making it harder for an attacker to gain information about
secure applications running over the NoC. Side-channel attackers
monitor physical, non-functional characteristics of a NoC
implementation (e.g. timing, power dissipation, temperature,
electromagnetic emissions), and try to identify patterns that can be
correlated with functional characteristics of a secure application
(e.g. time to decrypt a message, length of a private key, location of
a private key on the chip's distributed
memory~\cite{ESL15}). 

This follows a commonly used approach based on the randomisation of the information leaking over side channels. There are
many of the components that are not easy to randomise unless one has
full control over the NoC design, such as the arbitration and flow
control latencies (which are defined by the NoC architecture) and the
physical characteristics of the NoC (which are defined by the chip
fabrication process). We consider those approaches to be outside the
scope of this paper and focus instead on approaches that do not
require significant changes on the NoC. Specifically, we focus on the
randomisation of packet routes. By randomly changing the route of
every packet injected into the NoC, we can introduce random effects to
all side-channels of interest, such as packet timing, energy dissipation, temperature and electromagnetic emissions. In this paper, we concentrate on a threat model based on packet timing, as described in the next subsection.

\subsection{Threat Model}\label{ThrMod}

In this paper, we assume that the NoC and its interfaces to the cores are secure. We also assume that secure tasks execute in secure cores (i.e. cores that do not allow the execution of unsecured tasks). For this threat model, we assume that the NoC communicates sensitive information between two secure tasks, which we refer as the sensitive communication. We then assume an adversary that has knowledge about the NoC architecture, about the mapping of secure tasks to (secure) NoC cores, and is able to gain control of at most two non-secure NoC cores. 

A successfull attack happens when the adversary is able to infect two cores that can communicate over a route that intersects with that of the sensitive communication. In that case, the adversary is able to use one of the infected cores to inject low priority packets into the NoC towards the second infected core. The latency interference imposed by the sensitive communication over the malicious low priority traffic can provide the attacker with valuable information about the timing, frequency and volume of the secure communication. 

This threat model is not new, and its variations have also been used in best-effort NoC-based systems by ~\cite{WAN12} and~\cite{FLO16}. The timing nature of the threat is also the same used in hard real-time uniprocessor systems by~\cite{Yoon16}.

By using a route randomisation approach, it is possible to prevent the adversary from obtaining accurate information about the sensitive communication. Because not every packet of the secure communication will interfere on the malicious flows injected by the attacker, the information about timing, frequency and volume they can obtain will be less accurate, which as a consequence increases the resilience of the NoC against the threat. There are many ways to introduce route randomisation in NoCs, and we will discuss our design decisions in subsection \ref{Design}. 

\begin{figure}[!htb]
  \centering
  \includegraphics*[width=\linewidth]{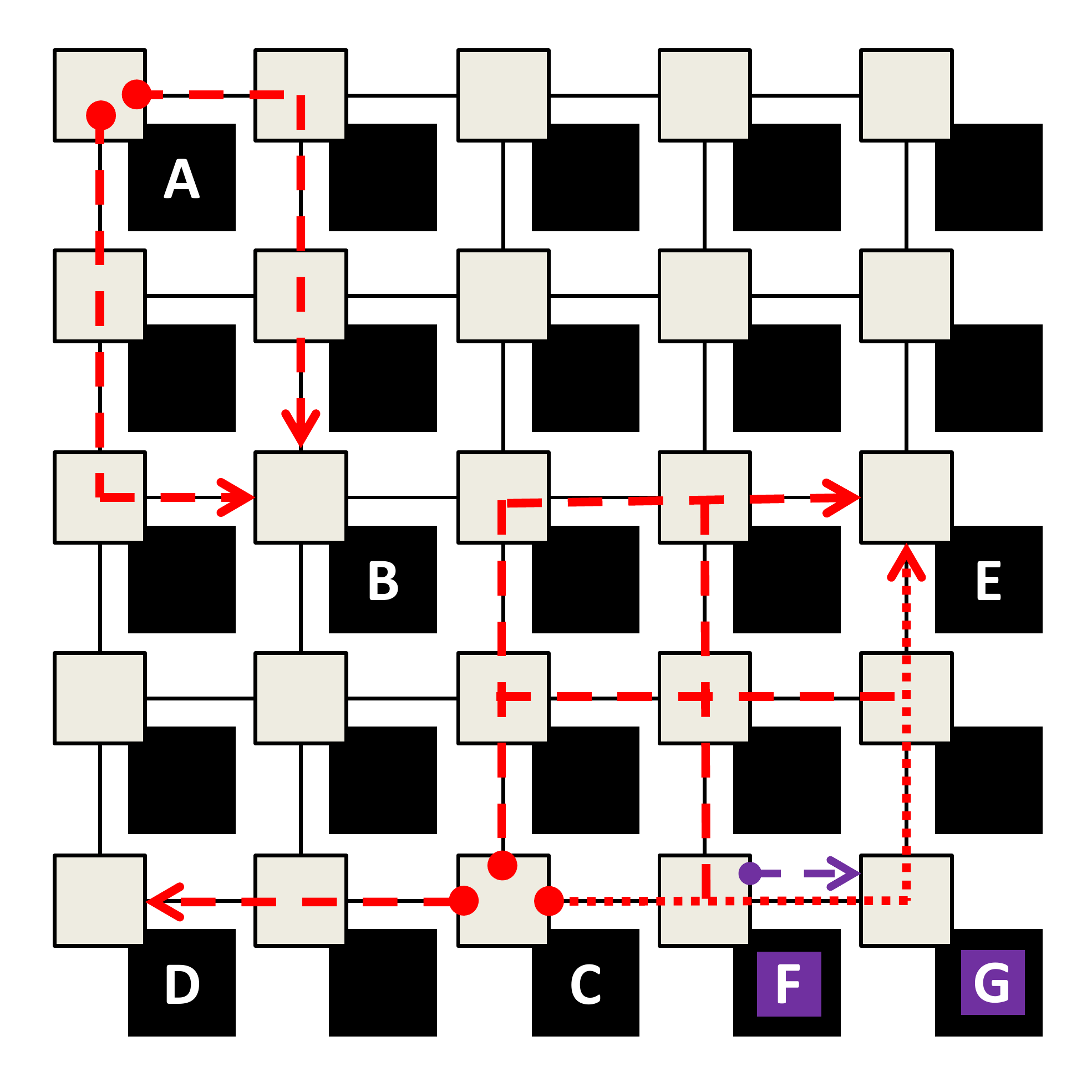}
  \caption{Threat model, and examples of route randomisation with pseudo-adaptive XY (from A to B) and west-first (from C to D and C to E) algorithms}
  \label{FIG-Routes}
\end{figure}

Figure \ref{FIG-Routes} shows an example of the described threat model. It shows an adversary controlling cores F and G, and using a malicious packet flow (shown as a purple dashed line) to infer data about a sensitive communication between secure cores C and E (shown as a red dotted line, representing the case of a NoC with deterministic XY routing). In the case of a NoC with randomised routing, all routes between C and E will be used (red dashed and dotted lines), preventing the adversary from inspecting the complete sensitive communication.

\subsection{System Model}\label{SysMod}

To increase NoC resilience against side-channel attacks while providing hard real-time guarantees to the
application tasks running on it, we must make assumptions about the application behaviour such as
upper-bounds on resource usage by every application task and packet. In this
paper, we follow the well-known and widely used sporadic task model,
which makes assumptions about the worst-case execution time (WCET) of
all tasks and their shortest inter-arrival interval (i.e. their
period). Since we are concerned about NoC communications, we follow an
extension of the sporadic task model that considers that tasks inject
packets to the NoC only after their execution completes, and that the
maximum packet size is known~\cite{Indrusiak14}.

Thus, a hard real-time application $\Gamma$ comprises $n$ real-time
tasks $\Gamma$ =$ \{\tau_1,\tau_2, \ldots, \tau_n\}$. Each task
$\tau_i$ is a 6-tuple $\tau_i$ = ($C_i$, $T_i$, $D_i$, $J_i$, $P_i$,
$\{\phi_i\}$) indicating respectively its worst case computation time,
period, deadline, release jitter and priority. The sixth element of
the tuple is an extension to the sporadic task model proposed
by~\cite{Indrusiak14}, and represents the communication packets sent
by $\tau_i$ at the end of its execution. Each packet $\phi_i$ is
defined as a 3-tuple $\phi_i$ = ($\tau_d$,$Z_i$,$K_i$) representing
its destination task, size and maximum release jitter. In this paper,
we assume for simplicity that a single packet is released at the end
of each execution of each task, but the contributions presented here
can be generalised for any number of released packets.

Such applications are executed over a NoC platform like the one
described in subsection \ref{NoC} above. We model such a platform as a
set of cores $\Pi$ =$\{\pi_a,\pi_b, \ldots, \pi_z\}$, a set of
switches $\Xi$ =$\{\xi_1,\xi_2, \ldots, \xi_m\}$, and a set of
unidirectional links $\Lambda$ =$\{\lambda_{a1},\lambda_{1a},
\lambda_{12}, \lambda_{21}, \ldots, \lambda_{zm}, \lambda_{mz}\}$. We
also model the mapping of tasks to cores with the function
$map(\tau_i)$ = $\pi_a$.

The routing of packets over the NoC can be modelled by the function
$route(\pi_a,\pi_b)$ = $\{\lambda_{a1},\lambda_{12},\ldots,\lambda_{mb}\}$, denoting the subset of $\Lambda$ used to transfer
packets from core $\pi_a$ to core $\pi_b$. We can then extend the
function $map$ to also model the mapping of a packet to its route:
$map(\phi_i)$ = $route(map(\tau_i), map(\tau_d))$.

With the knowledge of the NoC architectural characteristics such as
the latency to cross a link or to route a packet header, and with the
knowledge of the length of a packet's route (i.e. its hop count, or
$|route(\pi_a,\pi_b)|$ as expressed in~\cite{Indrusiak14}), it is
possible to calculate the no-load latency $L_i$ of every packet
$\phi_i$: the time it takes to completely cross the NoC from its
source to destination without any interference or contention from
other packets. For the NoC described in subsection \ref{NoC}, and for
most commercial and academic NoCs, the no-load latency of a packet can
be deterministically obtained, and will not change if its route and
the NoC operation frequency do not change.

\section{NoC Routing Randomisation}\label{Random}

\subsection{Design Choices and Constraints}\label{Design}

There are many design choices related to packet routing in different NoC architectures~\cite{Pasricha10}. As expected, those choices also define whether and how route randomisation can be achieved. For example, some NoC architectures use deterministic routing~\cite{Moraes04}, meaning that there is only one possible route between a source and a destination, effectively preventing the approach proposed here. Among NoCs supporting dynamic or adaptive routing, which are the ones we target, there is a key design choice affecting the randomisation approach: source or distributed routing. 

In source-routed NoCs, the routing decision is done by the source core or its respective NI. This is usually implemented as multiple packet header flits that contain the next-hop information for each of the switches along the packet's route. Once a switch routes one of the packet headers by assigning its output port, it discards that header flit and forwards the rest of the packet through that port. The next switch will route the subsequent header flit, discard it, forward the rest of the packet, and this is repeated all the way towards the packet destination. By following this approach, it is possible to program the source core or its NI to perform full route randomisation before every packet release. 

In NoCs with distributed routing, the next-hop decision is made by each switch individually. Typically, they have far less resources than the cores (and often than the NIs), so the routing decisions are based on simple rules related to the relative position of the destination core with regards to the switch holding the packet header (e.g. pseudo-adaptive XY~\cite{Dehyadgari05}, turn model~\cite{Glass92}). In those cases, it is only possible to randomly choose from a predefined subset of all possible routes. For instance, pseudo-adaptive XY switches can only randomly choose between two routes between a source and a destination (e.g. routes between cores A and B in Figure \ref{FIG-Routes}). Switches implementing turn model routing may have a larger number of alternative routes to randomly choose from in most cases, but must behave deterministically for some specific cases. Figure \ref{FIG-Routes} shows two routes created by a west-first turn model: packets between core C and D have only one possible route, as the destination is located on the west of the source, while packets from core C to E can take a variety of possible routes.

In both source and distributed routing, the NoC component making
random decisions must have access to a source of random data, such as
a pseudo-random number generator (PRNG, generated by a deterministic
algorithm) or a true random number generator (TRNG, often generated
out of low level noise signals). Such sources can have significant
hardware overhead, thus favouring source routing because of the low
area constraints for NoC switches. For the route randomisation
approaches reviewed above, however, overheads should be minimal in
either case as they only require random sources with one-bit output.

Additional issues when randomising packet routes include the potential
increase of the packet route, the possibility of deadlocks, and the
potential increase of packet latency (and therefore the potential
violation of real-time constraints). Let us now address each of them.

All the routing approaches reviewed above are minimal: the route they
choose has the smallest possible hop count between source and
destination. This is because of their obvious advantages in terms of
latency, network contention and energy dissipation. However, from the
point of view of side-channel attack resilience, it may be interesting
to exploit non-minimal randomised routing in order to decorrelate the
side channels with the functional properties of the packet
communication (e.g. short packet transmission between neighbouring
cores would not necessarily have the shortest latency and lowest
energy dissipation if they are forced to take a long route across the
chip).

Deadlock-free packet communication is a critical characteristic for
NoCs. This can be achieved at the link arbitration layer, e.g. with
priority-preemptive virtual channels~\cite{Indrusiak14}, or at the
network layer by restricting the possible turns of the routing
algorithm (either in source or in distributed routing). In NoCs that
ensure deadlock-freeness at the network layer, special care must be
taken by the route randomisation approach to avoid introducing turns
that can lead to deadlocks.

Finally, route randomisation is likely to change the latencies of
packets, both because for every release their routes may have
different hop counts (leading to different no-load latencies) and
because different routes may trigger different contention scenarios
(leading to different blocking times). In our approach, such
variability is actually desirable because it is a key aspect to
increasing the NoC's resilience against side channel attacks. In the
case of hard real-time systems, however, it is critical that such
variability is bounded and that the worst-case latencies of all
packets are always less than their deadlines. In the next subsection,
we propose an extension to existing schedulability analysis to
evaluate if that is the case for a given application mapped to a given
NoC architecture. The proposed approach is simple, yet general enough
to analyse randomised routing approaches following any of the design
choices reviewed above: source or distributed, minimal or non-minimal,
and with deadline-freeness ensured at the link or network layer.

\subsection{Schedulability Analysis}\label{Sched}

Schedulability analysis for a set of sporadic packets transferred over
a priority-preemptive wormhole switching NoC was presented
in~\cite{Shi08}. A set of packets is deemed schedulable if the
worst-case latency of each packet is less than their deadline. By
coupling that analysis with classical response time analysis for
uniprocessor fixed-priority scheduling, an end-to-end schedulability
analysis for that type of NoC was proposed in~\cite{Indrusiak14},
considering the worst-case response times of tasks and the worst-case
latency of the packets they generate. Both the original analysis
from~\cite{Shi08} and the end-to-end extension from~\cite{Indrusiak14}
assume static routing, so a different formulation is needed before it
can be used for the purpose of this paper. First, we review those
formulations, but using the notation described in subsection
\ref{SysMod}.

According to \cite{Shi08}, the worst-case latency $S_i$ of a packet
$\phi_i$ can be obtained from Equation \ref{stan}. This equation is
defined recursively and iterated until a stable fixed point is
discovered.
\begin{equation}\label{stan}
S_i \ = \ L_i \ + \  \sum_{\phi_j \in \mathbf{interf}(i)} { \left\lceil
{\frac{S_i + K_j + K_j^I}{T_j}} \right\rceil L_j },
\end{equation}
The set $\mathbf{interf}(i)$ is the set of higher priority packets
$\phi_j$ whose route shares at least one link with the route of
$\phi_i$ and therefore can interfere with it. Precisely,
$\mathbf{interf}(i)$ = $\{ \phi_j \in \phi : map(\phi_i) \cap
map(\phi_j) \ne \emptyset \}$. The two terms $K_j$ and $K_j^I$ denote
respectively the maximum release jitter of the interfering packet
$\phi_j$ and its maximum indirect interference jitter. As shown
in~\cite{Indrusiak14}, $K_j$ is equal to the worst case response time
$R_j$ of task $\tau_j$ which produces $\phi_j$, assuming that $\phi_j$
will be released immediately after the end of $\tau_j$'s
execution. $R_j$ can be calculated using uniprocessor response time
analysis, considering the type of task scheduling by the operating
system at each core (e.g. priority-preemptive). And as shown in
\cite{Shi08}, the indirect interference jitter $K_j^I$ can be bound by
$S_j - L_j$.

It can be seen in Equation \ref{stan} that the route of a packet
affects its worst-case latency because it defines the set of packets
that can add to the interference term of the equation (i.e. sum
operator). Route randomisation would change the set
$\mathbf{interf}(i)$ at each packet release, since different routes
would produce different interference patterns. An intuitive way to
find the worst-case latency of a packet with a randomised route would
be to calculate the worst-case latency of each of its possible routes
with Equation \ref{stan}, and pick the highest value. However, that
approach works only if there is a single packet with randomised route,
and all others following deterministic routes.

A general analysis where all packets could potentially have randomised
routes is more complex: all possible routes of a packet would have to
be tested with all possible routes of all other packets before the
worst case could be found. Furthermore, if one cannot make
probabilistic assumptions on the randomisation approach, pathological
cases must also be taken into account (e.g. the same route could be
chosen again and again for a single packet over a long period of time,
even though that is very unlikely).

In this paper we assume that, in the worst case, if there is a way for
a high-priority packet to interfere with a low priority packet, it
would interfere with it in every possible release. This means that
even though there may be routes when packets do not interfere with
each other, we assume that in the worst case the random choice of
route would always pick the ones where there is interference. This is
perfectly reasonable when packets have similar periods, but it gets
more and more pessimistic as we reduce the periods of higher priority
packets. In that case, high priority packets would have a larger
number of releases within a single release of a low priority packet,
thus interfering more often with it, even though the larger number of
releases would make less likely that an interfering route would be
chosen every time.

To calculate worst-case latencies for the general problem where all
packets could have randomised routes, we define the set
$\mathbf{interf_r}(i)$ as the set of higher priority packets $\phi_j$
who could, with any of their possible routes, interfere with any of
the possible routes of the packet of interest $\phi_i$. To precisely
define that set, we must first define a new function
$route_r(\pi_a,\pi_b)$ =
$\{\lambda_{a1},\lambda_{12},\lambda_{13},\lambda_{14},
\dotsc,\lambda_{mb}\}$, denoting the subset of $\Lambda$ that contains
all the links that could be part of any of the routes that could be
randomly chosen to transfer packets from core $\pi_a$ to core $\pi_b$,
and a new function $map_r(\phi_i)$ = $route_r(map(\tau_i), map(\tau_d))$. Then,
$\mathbf{interf_r}(i)$ = $\{ \phi_j \in \phi : map_r(\phi_i) \cap
map_r(\phi_j) \ne \emptyset \}$.

By applying Equation \ref{stan} with the summation over the set
$\mathbf{interf_r}(i)$ instead of the original $\mathbf{interf}(i)$,
we can then find an upper bound to the packet latencies over a NoC with
randomised routing.

\subsection{Optimising the Performance-Security Trade-off}\label{Opt}

The schedulability analysis proposed in the previous subsection can
only be used to test whether a particular randomised NoC configuration
can meet the hard real-time constraints of an application. It offers
no alternatives in case of negative results, i.e. when performance
constraints are not met. In this subsection we show how the
schedulability test can be exploited as a fitness function in a design
space exploration process. Similarly to~\cite{Sayuti13}
and~\cite{Indrusiak14}, we follow an evolutionary approach to navigate
over a key part of the design space: task-core mapping. By changing
that mapping, it is possible to achieve fine-grained improvements on
schedulability of tasks over cores and packet flows over NoC
infrastructure (e.g. tasks that are barely unschedulable can become
schedulable by a simple remapping of one of the higher priority tasks
that interfere with their computation or communication, thus changing
the set $\mathbf{interf}$ in Equation \ref{stan}). The same can
happen in the case of route randomisation, since changes on mapping
can determine which randomised routes interfere with each other and in
turn affect schedulability through changes in the $\mathbf{interf_r}$
set.

\begin{figure}[!htb]
  \centering
  \includegraphics*[width=\linewidth]{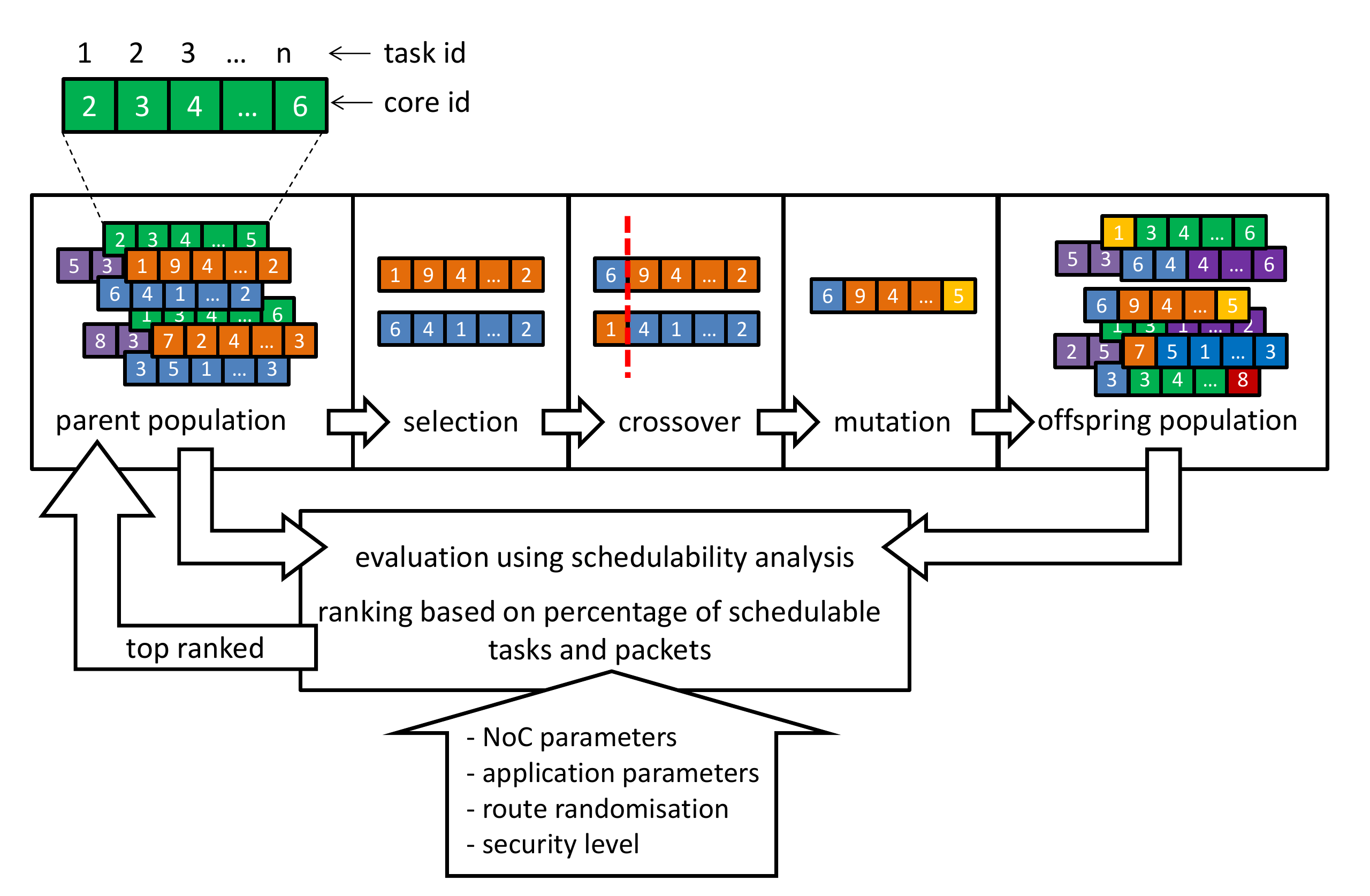}
  \caption{Evolutionary pipeline to optimise performance-security trade-off}
  \label{FIG-GA}
\end{figure}

Figure \ref{FIG-GA} shows the evolutionary pipeline proposed here,
which start with an arbitrary population of task mappings using a
given route randomisation approach and a given level of security. It
then uses evolutionary operators such as mutation and crossover to
improve the mapping population with regards to the percentage of
schedulable tasks and packets calculated using the proposed
modification of Equation \ref{stan}. For every generation of the
population, those with the larger number of schedulable tasks and
packets are selected to the next generation, where they will be again
mutated, crossed-over, evaluated and selected to the subsequent
generation. The pipeline stops after a fully schedulable mapping is
found, or a predefined maximum number of generations is reached.

Unlike many constructive task mapping approaches, the evolutionary pipeline proposed here does not necessarily try to map communicating tasks to the same or neighbouring cores. Its fitness function can be tuned, for instance, to keep communicating tasks as far apart as possible while keeping their communication packets schedulable over a variety of randomly-chosen routes.

In this paper, we consider two types of route randomisation which can
be implemented either as source or distributed routing, namely random
XY/YX and random west-first. Random XY/YX is a randomised version of
pseudo-adaptive XY routing used in~\cite{Dehyadgari05}, so the route
of the packet to its destination is randomly chosen between the XY or
the YX route prior to the injection of the packet header into the
network. In random west-first, we randomise one of the turn model
routing approaches~\cite{Glass92} so that whenever a packet is allowed
more than one route it randomly chooses one of them (i.e. uniform
probability among all alternatives).

We then allow for multiple levels of security by changing how many
packet flows are allowed to have their routes randomised. A baseline
with no randomisation should have the best results regarding
schedulability, given that packets suffer less interference and
therefore are more likely to be schedulable. Then, increased levels of
security can be achieved by randomised larger percentages of packet
flows, up to a fully randomised configuration where all packets follow
randomised routes on every release. In the next section, we show
experimentally that the proposed schedulability test and evolutionary
optimisation pipeline can produce NoC configurations able to hold hard
real-time guarantees with maximised security potential.

\section{Experimental Work}\label{Eval}

We evaluate the proposed approach in two distinct experimental
setups. The first uses the proposed schedulability test and
evolutionary pipeline to balance the trade-off between performance
guarantees and security over a large set of synthetically generated
applications. The second uses a cycle-accurate NoC simulator to show
the effects of route randomisation upon latency with a realistic
application.

\subsection{Schedulability-driven optimisation of route randomisation}\label{Eval-sched}

This section presents the workflow for analytic schedulability
evaluation, and evolution with an evolutionary pipeline based on a
genetic algorithm (GA). It follows the pipeline presented in Figure
\ref{FIG-GA}. To evaluate the challenge of optimising different
applications with different levels of load, we synthetically generate
thousands of applications, each of them composed of tasks that
communicate with each other with different numbers of packet flows. We
then apply the evolutionary pipeline to each one of those
applications, aiming to optimise the mappings of tasks in such a way
that the whole set of tasks and flows is schedulable at different
levels of security. We then plot the percentage of schedulable
applications we could achieve for each level of security and each
level of load. For the sake of reproducibility, we provide below more
details on the whole process.

For a single experiment upon a given NoC and set of parameters
(e.g. topology, operating frequency, switch and link latencies), a
range of packet flow counts are identified, each of which represents a
level of communication within the application, and therefore a
utilisation load upon the NoC. For each flow count chosen for
experimental evaluation, a set of tasksets and packet flowsets are
generated, each containing the chosen number of flows. The number of
tasks is kept roughly constant, and all of them are either source or
destination of at least one packet flow. Therefore, flowsets with
higher flow counts represent increasing packet contention between the
same endpoints. Flows are assigned to particular source and
destination tasks with uniform random probability. This implies that
the average number of flows transmitted is even across all tasks,
although as a result of the random assignment there may be unique
hotspots.

Following this, an experiment is initialised by defining a population
of initial mappings, and a setting for the target level of security
case setting. The levels of security settings are defined as either
unsecured, or 25\%, 50\%, 75\% and 100\% secured flows. The secured
flows are those that will use randomised routing, providing increased
potential protection against side-channel attacks. In case of a
partial provision of security e.g. 50\%, security is assigned to the
flows in their order of priority, with the highest priority flows
being randomised. The rationale is to enforce overall random
interference patterns, since higher priority packets are the ones
causing interference.

A population of chromosomes (each representing of a mapping of tasks
to cores upon the NoC, as shown in the upper-left corner of Figure
\ref{FIG-GA}) is specified for each level of load (i.e. synthetically
generated taskset and flowset with a specific flow count). A genetic
algorithm is then used to evolve these chromosomes, performing
mutation, crossover and evaluation of the population according to a
fitness function based on the modified Equation \ref{stan}. This is
done separately for each level of security, each of them generating a
different $\mathbf{interf_r}(i)$ set representing the randomised
routes of different packet flows.

By applying the modified Equation \ref{stan} for every packet flow
of the application, it is possible to check whether each of them is
schedulable, i.e. their end-to-end latency is less than the respective
deadline. The overall fitness of an application is then assumed to be
the number of schedulable packet flows. Following the fitness function
evaluation, the population is culled to retain only the chromosomes
that are at the top of the fitness ranking. If the fitness function
indicates that the top-ranked chromosome represents a mapping where
all flows are schedulable, then the GA terminates early. Otherwise,
following the completion of the chromosome improvement process at a
fixed number of generations, the best chromosome (output mapping) and
schedulability obtained (both aggregate flows and flowsets) is output
for display.

To show the impact of the level of security on performance guarantees and resource usage, we have produced several experimental series:
\begin{LaTeXdescription}
\item[No security (NS)] Deterministic routing, fitness function incorporates schedulability
  calculated using Equation \ref{stan} with the original $\mathbf{interf}(i)$ set.
  
\item[Percentage security (PS(\%))]A given percentage of the packet
  flows use randomised routing, fitness function evaluated using
  Equation \ref{stan} with the proposed $\mathbf{interf_r}(i)$ set
  reflecting that percentage.
  
\item[Application of security a posteriori (SAP)]Evolution is
  performed using a fitness function that tests the schedulability
  without any security mechanisms (only deterministic routing), aiming
  to find a schedulable mapping without security
  considerations. Following the completion of this evolutionary
  process, the evolved best application mapping has 100\% of its
  packet routes randomised, and is then evaluated with Equation
  \ref{stan} with the proposed $\mathbf{interf_r}(i)$ set. This
  experiment therefore aims to show that the optimisation of the
  mapping should take into account route randomisation, and that poor
  results can be expected from applying randomisation to a mapping
  that was optimised for deterministic routing.
\end{LaTeXdescription}

\begin{table}[!tb]
  \begin{tabular}{|l|l|l|}
    \hline
    NoC/Packet flowset parameters & Value\\
    \hline
    Maximum packet flow no-load latency & 100 ms\\
    Maximum period & 500 ms\\
    Priority assignment & Deadline monotonic\\
    Route randomisation & Random XY/YX\\
    Standard NoC topology   & 4x4\\
    Enlarged NoC topology & 8x8\\
    Flowsets per data point & 100\\
    \hline
    GA parameters & \\
    \hline
    Population size & 100\\
    Mutation individual task moving probability & 0.3\\
    Maximum generations & 50\\
    \hline
  \end{tabular}
  \caption{Evaluation parameters}
  \label{TAB-SIM-PARAMS}
\end{table}

\subsubsection{Results}

\begin{figure}[!tb]
  \centering
  \includegraphics*[width=\linewidth,height=5.5cm]{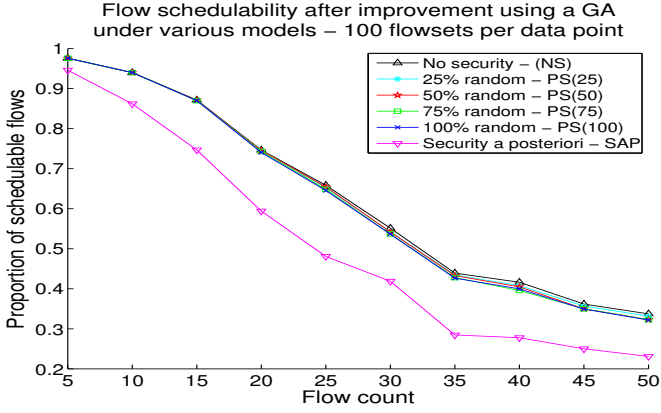}
  \caption{Flow schedulability results in the 4x4 case}
  \label{FIG-SCHED-FLOWS-4x4}
\end{figure}

\begin{figure}[!tb]
  \centering
  \includegraphics*[width=\linewidth,height=5.5cm]{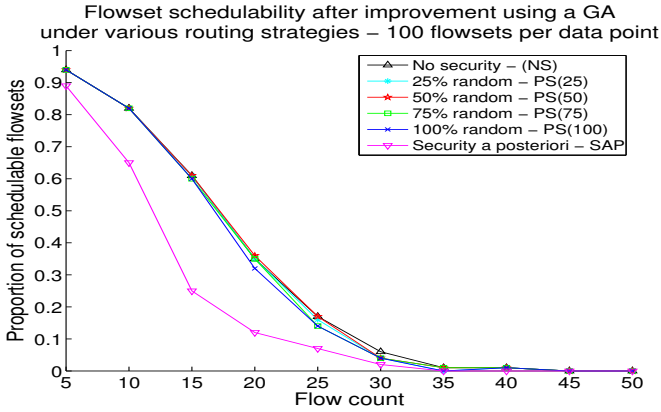}
  \caption{Flowset schedulability results in the 4x4 case}
  \label{FIG-SCHED-FLOWSETS-4x4}
  \vspace{-0.3cm}
\end{figure}

Figure \ref{FIG-SCHED-FLOWS-4x4} shows the aggregate schedulability of
flows after improvement with the GA, as a mean proportion across all
flowsets generated for that data point. It is clear that the ordering
of the results series in the illustrated plot follows the proportion
of security provided, with an increasing number of flows in the
flowsets (and therefore an increasing load upon the NoC) providing a
slight reduction in schedulability of the evolved cases. This is as
anticipated, in that the worst-case schedulability analysis would be
affected by the increased interference present from the optional
random routes. However, since each GA run is an independent
evolutionary process, the ordering of the series does not always
follow the anticipated order. In the SAP series (security a
posteriori), evolution is performed using a fitness function that
tested schedulability under the no security case (XY
routing). However, following the completion of the GA the evolved
mapping schedulability was evaluated with all flows using randomised
routing. As anticipated, the schedulability of SAP is considerably
worse than the NS or PS series, since the evolution was performed
using a routing strategy that assumes lower interference than the
final evaluation case. Figure \ref{FIG-SCHED-FLOWSETS-4x4} shows the
schedulability of flowsets. A flowset is only considered schedulable
if every flow within it is schedulable. The results follow the same
general trend as in Figure \ref{FIG-SCHED-FLOWS-4x4}, although they
reach zero earlier since flowset schedulability requires every
component flow to be schedulable.

\begin{figure}[!tb]
  \centering
  \includegraphics*[width=\linewidth,height=5.5cm]{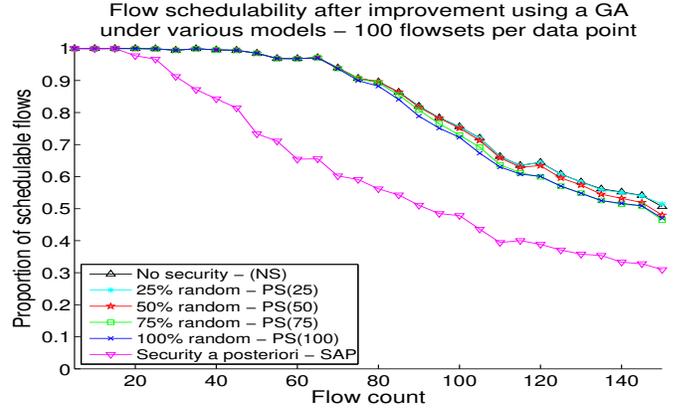}
  \caption{Flow schedulability results in the 8x8 case}
  \label{FIG-SCHED-FLOWS-8x8}
\end{figure}

\begin{figure}[!tb]
  \centering
  \includegraphics*[width=\linewidth,height=5.5cm]{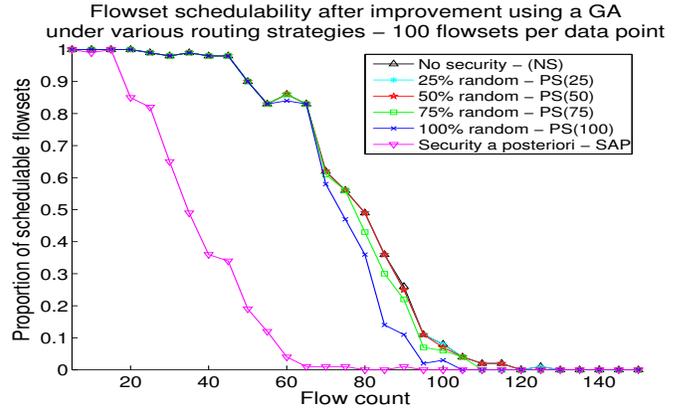}
  \caption{Flowset schedulability results in the 8x8 case}
  \label{FIG-SCHED-FLOWSETS-8x8}
  \vspace{-0.3cm}
\end{figure}

For the 8x8 example evaluation case, the results are presented in
Figures \ref{FIG-SCHED-FLOWS-8x8} and
\ref{FIG-SCHED-FLOWSETS-8x8}. The results show a greater separation
between the NS and PS series after NoC evolution, due to the increased
NoC size and number of flows allowing a greater complexity of
interference graphs when randomised routing is enabled. The SAP case
also has significantly lower schedulability, since its evolved mapping
was obtained without routing randomisation and imposing randomisation
later affects schedulability. In the schedulability of flowsets in
Figure \ref{FIG-SCHED-FLOWSETS-8x8}, it is clear there is a wider
difference in schedulability between the PS(100) secured case and NS
(no security) particularly in flowsets with 70 to 85 flows.
This illustrates that as the interference graph becomes more complex it is
harder for the GA to find schedulable mappings.

\subsection{Cycle-accurate simulation of route randomisation}\label{Eval-sim}
One of the key concerns in altering network routing is the impact that
it will have upon latency for packet transmission, particularly in
latency-sensitive real time applications. This section considers via
simulation the impact of randomising of the routing protocol on the
latency of a previously published real-time application case, the
autonomous vehicle application \cite{Indrusiak14}.

The simulation framework used for this section is a cycle-accurate NoC
model with support for priority preemption and virtual channels. This
simulator has been extensively validated in our previous work,
frequently being used as a baseline for results in latency and power
analysis~\cite{Indrusiak15}~\cite{Harbin16}.

\subsubsection{Application Structure}
The application used in this application is an autonomous vehicle (AV)
application~\cite{Indrusiak14}. This application consists of 38
communicating flows between a set of tasks that represent video
processing, system monitoring and control for a robotic vehicle. As is
the convention throughout this paper, priorities are defined such that
lower priority index values represent the highest priority
transmissions. The priorities, data transmission rates, frequencies
and deadlines of these application transmissions are as defined in
\cite{Indrusiak14}, although a different mapping has been used in
order to show the impact of routing protocols on a randomly selected
mapping without artificial tuning to favour a particular routing
protocol. The application has been mapped onto a 4x3 NoC, and the
video resolution of the AV application video streams is 640x480. Since
the application mapping is static and a single priority level is used
per packet, a packet always travels between a fixed source-destination
pair during the simulation.

\subsubsection{Routing Alternatives}
In this simulation evaluation, two routing alternatives incorporating
randomisation are used, in addition to the baseline comparison of XY
routing. The first routing alternative uses the XY/YX approach. In
this approach, traffic producers determine uniformly randomly on
injection whether a data packet will use XY or YX routing, and
following this decision a flag is set in the data packet to control
the routing behaviour. As a result, the chosen routing algorithm
(either XY or YX) is used throughout packet transmission.

In addition, an alternative routing structure known as random west
first (RWF) routing is also implemented, which allows randomised
routing decisions to be taken by individual arbiters during data
transmission. RWF requires the packet always be forwarded towards the
west when the destination node is west of the current
arbiter. However, any other destination port can be chosen uniformly
randomly (east, north or south) as long as the direction taken is
towards the destination. Therefore, the RWF approach permits a more
diverse range of transmission paths than the XY/YX selection approach,
providing more potential protection against side channel attacks.

\subsubsection{Evaluation Results}
\begin{figure}[!tb]
  \centering
  \includegraphics[width=1.0\linewidth]{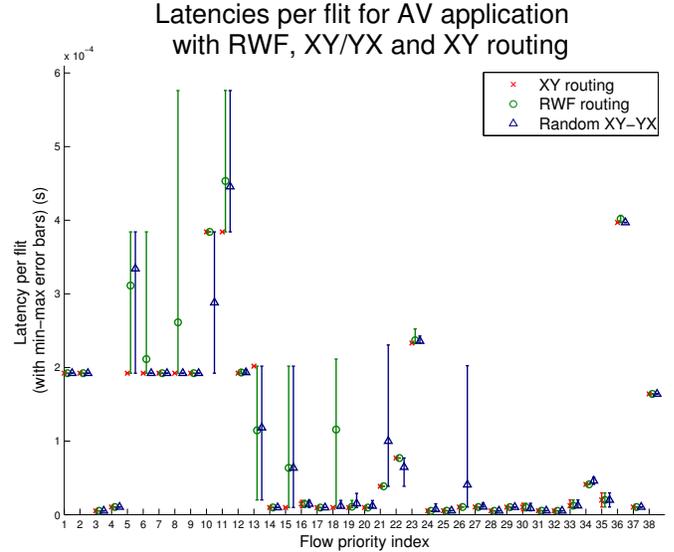}
  \caption{Communication latency results for the randomised routing case on the AV application}
  \label{FIG-SIM-RESULTS}
\end{figure}

\begin{figure}[!tb]
  \centering
  \includegraphics[width=1.0\linewidth]{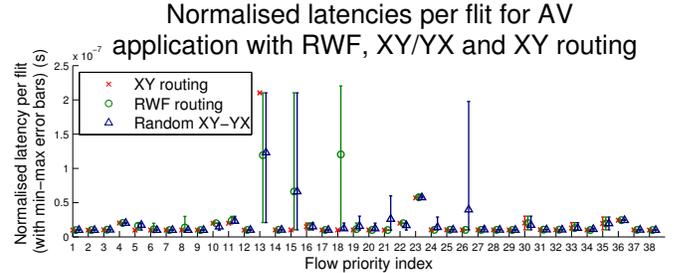}
  \caption{Communication latency results (normalised) for the randomised routing case on the AV application}
  \label{FIG-SIM-RESULTS-NORM}
\end{figure}

The results are presented in Figures \ref{FIG-SIM-RESULTS} and
\ref{FIG-SIM-RESULTS-NORM}, illustrating the max-min-mean latencies
and normalised latencies for the randomised routing cases (XY/YX and
RWF) versus the baseline. Normalised latency is calculated by dividing
the end-to-end latency of the packets by the packet size, which
provides a metric of latency per flit. This metric is therefore more
sensitive to delays in the transmission of short packets.

The latency results presented in Figure \ref{FIG-SIM-RESULTS}
illustrate that routing randomisation typically increases the
communication latencies for the majority of packets compared to fixed
XY routing. This is particularly evident in the case of the packets
with priority 8 under RWF routing, which experience an increased
latency due to contention with other higher priority flows on some of
the randomly chosen routes. In the XY/YX routing case, increased
latency is also observed for the packets with priorities 21 and 26 in
some cases. Interestingly, for some of the packet transmissions with
priority 10 and 13, the use of randomised routing is also to reduce
latency in the best case, either by routing a higher priority packet
so that it no longer causes interference, or routing the current
packet around the interferer.

Considering the normalised latency results in Figure
\ref{FIG-SIM-RESULTS-NORM}, it is clear that the relative impact of
route randomisation is most significant upon packets with priorities
13, 15, 18 and 26. These transmissions represent some of the shortest
packets in the system, which are therefore more greatly impacted on a
relative basis by contention with other packets. As depicted in the
previous figure, some priority 13 packets encounter a large reduction
in latency during some transmissions as a result of avoiding
interference.

\section{Conclusions and Future Work}\label{conc}

This paper has addressed the trade-off between security and hard
real-time performance guarantees in Networks-on-Chip. It has proposed
route randomisation as a way to increase NoC resilience against
side-channel attacks, and has discussed a number of design
alternatives for the randomisation approach. It then has proposed a
schedulability test for applications running over a secure
priority-preemptive NoCs using route randomisation. Finally, the paper
identifies an optimisation pipeline which can be guided by the
proposed schedulability test towards configurations that can achieve
full schedulability while maximising the provided level of
security. Extensive experimental work using 4x4 and 8x8 NoCs with
random XY/YX routing running thousands of synthetically generated
applications show the performance guarantees that can be achieved by
the proposed approach at four different levels of security, compared
against two baselines (no security, and full security applied a
posteriori). Additional experiments with a realistic application
running over 4x3 NoCs with random XY/YX and random west-first routing
were performed with a cycle-accurate simulator, aiming to show the
impact of route randomisation on latency variability, which in turn
shows the increased resilience against side-channel attacks.

Since this is the first paper addressing the trade-off between
security and hard real-time performance in NoCs, it had to make
several assumptions to be able to attack the problem. Lifting some of
those assumptions will certainly open new avenues of research, such as
using different NoC arbitration mechanisms (e.g. TDM) or different
route randomisation techniques (e.g. if randomised routes of
subsequent releases of packets are never the same, a less pessimistic
schedulability test can be used). Addressing those cases will require
new schedulability tests, but could still reuse the proposed
optimisation pipeline.

\subsection*{Acknowledgements}
The research described in this paper is funded, in part, by the
EPSRC grant, MCC (EP/K011626/1). No new primary data were created during this study.

\bibliography{refs}
\bibliographystyle{plain}
\end{document}